\documentclass[ twocolumn, showpacs,aps,superscriptaddress,prd,notitlepage,showkeys,nofootinbib]{revtex4-1}

\usepackage[normalem]{ulem}
\usepackage{amssymb}
\usepackage{amsmath}
\usepackage{graphicx}
\usepackage{dcolumn}
\usepackage{hyperref}
\hypersetup{colorlinks,urlcolor=blue,citecolor=blue,linkcolor=blue}
\usepackage{color,units}
\usepackage[dvipsnames]{xcolor} 
\usepackage{lineno}
\usepackage{xspace}
\usepackage{longtable} 
\usepackage{float}  
\usepackage{aas_macros}
\usepackage{amsfonts,wasysym,epsfig,verbatim,subfigure,bm,mathrsfs,lipsum}

\usepackage{color}
\usepackage{hyperref}
\usepackage{amsmath}
\usepackage{multirow}
\usepackage{graphicx}
\usepackage{envmath}
\usepackage{natbib}

\newcommand{\IUCAA}{Inter-University Centre for Astronomy and
  Astrophysics, Post Bag 4, Ganeshkhind, Pune 411 007, India}

\newcommand{\WSU}{Department of Physics \& Astronomy, Washington State University, 1245 Webster, Pullman, WA 99164-2814, U.S.A.}

\begin{document}

\title[]{Gravitational waves from binary black hole mergers surrounded by scalar field clouds: Numerical simulations and observational implications}

\author{Sunil Choudhary}
\affiliation{\IUCAA}

\author{Nicolas Sanchis-Gual} 
\affiliation{Centro de Astrof\'isica e Gravita\c{c}\~ao - CENTRA, Departamento de F\'isica, Instituto Superior T\'ecnico - IST, Universidade de Lisboa - UL, Avenida Rovisco Pais 1, 1049-001, Portugal}

\author{Anshu Gupta} 
\affiliation{\IUCAA}

\author{Juan~Carlos~Degollado} 
\affiliation{Instituto de Ciencias F\'isicas, Universidad Nacional Aut\'onoma de M\'exico,  M\'exico
Apdo. Postal 48-3, 62251, Cuernavaca, Morelos, M\'exico}

\author{Sukanta Bose}
\affiliation{\IUCAA} 
\affiliation{\WSU}

\author {Jos\'e A. Font}
\affiliation{Departamento de Astronom\'{\i}a y Astrof\'{\i}sica,
Dr. Moliner 50, 46100, Burjassot (Val\`encia), Spain}
\affiliation{Observatori Astron\`omic, Universitat de Val\`encia, C/ Catedr\'atico 
 Jos\'e Beltr\'an 2, 46980, Paterna (Val\`encia), Spain}

\begin{abstract}
We show how gravitational-wave observations of binary black hole (BBH) mergers can constrain the physical characteristics of a scalar field cloud parameterized by mass $\tilde{\mu}$
and strength $\phi_0$ that may surround them. We numerically study the inspiraling equal-mass, non-spinning BBH systems dressed in such clouds, focusing especially on the gravitational-wave signals emitted by their merger-ringdown phase. These waveforms clearly reveal that larger values of $\tilde{\mu}$ or $\phi_0$ cause bigger changes in the amplitude and frequency of the scalar-field-BBH ringdown signals. We show that the numerical waveforms of scalar-field-BBHs can be modelled as chirping sine-Gaussians, with matches in excess of 95\%. This observation enables one to employ
computationally expensive Bayesian studies for estimating the parameters of such binaries. Using our chirping sine-Gaussian signal model we  establish that observations of BBH mergers at a  distance of 450 Mpc will allow to distinguish BBHs without any scalar field from those with a field strength $\phi_0 \gtrsim 5.5\times 10^{-3}$, at any fixed value of $\tilde \mu \in [0.3,0.8]$, with 90\% confidence or better, in single detectors with Advanced LIGO/Virgo type sensitivities. This provides hope for the possibility of determining or constraining the mass of ultra-light bosons with gravitational-wave observations of BBH mergers.[This manuscript has been assigned the preprint number \textcolor{red}{LIGO-2000375}]
  
\end{abstract}

\maketitle

\section{Introduction}
\label{sec:intro}

The gravitational-wave (GW) detector network comprising the Advanced LIGO  (aLIGO) and the Advanced Virgo (AdV) interferometers recently launched the era of GW astronomy. The landmark observation in 2015 of a GW signal from a binary black hole (BBH)  merger and the subsequent additional detections of binary mergers that followed during the first two observational campaigns (O1 and O2), including a binary neutron star (BNS) system~\cite{GWTC-1}, have put GW astronomy on a very firm footing. Moreover, during O3 GW candidate events have been released as public alerts to facilitate the rapid identification  of  electromagnetic  or  neutrino  counterparts, expanding the capabilities of multi-messenger astronomy. A significant number of candidates have been publicly announced on the GW candidate event database~\footnote{{\tt gracedb.ligo.org}} and some confirmed detections have already been published~\cite{GW190412,GW190425,GW190814,GW190521,GW190521I}.

Accurate computations of the gravitational waveform of a compact binary coalescence event, especially in the early inspiralling stage, yield a plethora of information about the binary and the physics of its components. While large banks of waveforms are available for BBHs, and also for BNS mergers and BH-NS systems, relatively less information is available about possible departures from those waveforms if the binary components were exotic (yet physically plausible) compact objects, e.g.~boson stars, Proca stars, gravastars, fuzzballs or wormholes.  However, there are ongoing theoretical efforts to investigate them~\cite{2012LRR....15....6L,bh_in_fuzzy_dm2019, NRbeyondGR_CQGissue,decihertz_gwastro_imbh,Aneesh:2018hlp,cardoso2019LRR}. 
In particular, the merger of binary compact objects formed by fundamental bosonic fields has been explored
in several works, including head-on collisions and orbital mergers of boson stars, oscillatons, and Proca stars~\cite{palenzuela2007head,cardoso2016gravitational,brito2015accretion,brito2016interaction,helfer2018gravitational,bezares2017final,PhysRevD.96.104058,dietrich2018full,clough2018axion,Bezares:2018qwa,PhysRevD.99.024017}, 
The potential of GW astronomy for new discoveries might eventually shed light on the actual existence in Nature of those theoretical proposals.

Using GW observations as probes of new physics is challenging but they also provide a brand new experimental channel to try and find answers to the biggest unsolved problems in fundamental physics (see, e.g.~\cite{Barack:2018yly}). Some of those are the nature of dark matter and dark energy, the physics in the early Universe, and possible extensions  of the Standard Model. A well-known example for physics beyond the Standard Model is provided by ultra-light bosonic particles.
The masses of the ultra-light bosons of the string axiverse can range from $10^{-33}$ eV to $10^{-10}$ eV \citep{Arvanitaki:2010}. Even though their masses can be smaller than those of known particles, their existence is possible if the coupling to ordinary matter is very weak. 

In particular scalar fields surrounding supermassive BHs in galactic centers have been proposed as candidates for dark 
matter~\citep{Sahni:1999qe,Hu:2000ke,Matos:2000ng,Matos:2000ss,2017PhRvD..95d3541H}. This model assumes that dark matter is composed of bosonic particles that may condensate into macroscopic objects  around BHs. The justification for this proposal requires long-lived scalar-field configurations. Their dynamics and lifetime have been studied both in the linear regime~\citep{Witek:2012tr,Barranco1:2012qs,Barranco2:2013rua,Dolan:2012yt,CruzOsorio:2010qs} and with non-linear simulations in general relativity \citep{PhysRevD.91.043005, PhysRevD.92.083001, PhysRevD.94.043004}, providing convincing support to the proposal.
Specifically, the early work by \cite{Barranco2:2013rua} in the linearized regime showed that {\it massive} scalar fields surrounding stationary and {\it non-rotating} BHs could indeed form such quasi-bound states as a result of the presence of a potential well due to the mass term. These states decay at infinity and are characterized by a complex frequency whose real part represents the actual oscillation frequency while the imaginary part gives, depending on the sign, either their decay rate or the growth rate, if a mode is superradiantly unstable. 

The superradiant instability operates in {\it rotating} black holes (but see~\citep{PhysRevLett.116.141101} for an academic setup with a Reissner-Nordstr\"om BH) where bosonic waves scattered off the BH extract energy and angular momentum and increase the energy of the field through the classical process of superradiance~\cite{Detweiler:1980,2007PhRvD..76h4001D} (see also~\citep{Brito-review} and references therein). The nonlinear realisation of superradiance in Kerr BHs was  recently shown by~\citep{east2017superradiant,east2017superradiant2} employing a vector (Proca) bosonic field (see also~\cite{2014PhRvD..89f1503E,Okawa:2015}). The end-state of this process is the formation of {\it hairy} 
BHs, i.e.~Kerr BHs surrounded by either scalar or vector hair, in which the bosonic field is in equilibrium (i.e.~synchronized) with the BH~\citep{herdeiro2014kerr,Herdeiro:2015gia}. Recent works~\citep{Arvanitaki:2017,Brito:2017,Yang:2017lpm,Palomba:2019,Baumann2019, PhysRevD.101.063020,PhysRevD.101.024019,PhysRevD.99.104030,PhysRevD.99.084042,arxiv_article:2007.12793} have investigated possible observational signatures of the bosonic clouds through the detection of the nearly monochromatic GWs they emit, providing procedures to e.g.~constrain the QCD axion~\citep{Arvanitaki:2017}, probe ultra-light bosons in BBH inspirals through the analysis of resonant transitions between growing and decaying modes of the clouds~\citep{Baumann2019} and estimate upper limits for the detectability of ultra-light bosons through direct GW searches~\citep{Brito:2017,Palomba:2019}. Recent attempts \cite{arxiv_article:2007.12793} promote the use of third-generation ground-based GW detectors combined with the spaced-based LISA mission to increase the chances of detection using a multiband technique. We also note that the direct detection of bosonic fields in the form of bosonic stars has been recently proposed in connection with GW190521~\cite{bustillo2020ultra}.  Lastly, for its relevance to our setup and results, we highlight the work of~\cite{Yang:2017lpm} which investigated the joint evolution of intermediate-mass BBH surrounded by a shell of an axion-like scalar field of different strenghts, finding that the dynamics of the mergers can be modified by the presence of the environmental scalar field cloud, which also impacts the GW emission.

In this work we also investigate if GW measurements can probe the existence of bosonic clouds around BHs but we employ a different setup to that  used in most previous works. We study if the presence of a scalar field cloud might actually be established through its imprint on the GWs from BBH mergers. Our goal is to show through a combination of numerical-relativity simulations and Bayesian inference if the actual network of GW interferometers can measure the differences in the waveforms induced by the presence of scalar field clouds around the coalescing BHs. To this aim we parameterize the cloud  by its mass $\tilde{\mu}$ and its strength $\phi_0$. Our investigation reveals that it may actually be possible to observationally distinguish  BBHs without any scalar field from those with a field strength of order $\phi_0 \gtrsim 5.5\times 10^{-3}$, at any fixed value of $\tilde \mu \in [0.3,0.8]$, 
with 90\% confidence or better, in single detectors with aLIGO or AdV type sensitivity, up to distances of about 450 Mpc. At smaller distances ($\sim$100-200 Mpc) even less strong fields might be distinguishable. 

This paper is organized as follows: In section \ref{sec:basic equations} we briefly summarise our basic framework to study the dynamics of BBH mergers endowed with bosonic clouds.  
Section~\ref{sec:nr} describes the numerical setup, the initial data, and the results of the numerical simulations for varying scalar-field parameters. The measurement and estimation of these parameters through Bayesian inference are discussed in Section~\ref{sec:parma_estimates} and we close with Section~\ref{sec:conclusions} which presents our conclusions. Appendix~\ref{app:match} provides specific details on our numerical waveforms. Throughout the paper we use natural units, $c = G =\hbar  =1$. 

\section{Basic Framework} 
\label{sec:basic equations}

Our approach to model BBH mergers surrounded by a scalar field environment considers a massive complex scalar field interacting through gravity with the BHs. The system is governed by the Einstein-Klein-Gordon theory. Details of the formulation have been given in \cite{PhysRevD.91.043005, PhysRevD.89.104032} which we briefly summarise here. We consider a complex scalar field $\Phi$ minimally coupled to gravity described by the action

\begin{equation}
S = \int d^{4}x \sqrt{-g} \left( \frac{1}{16\pi}R - \frac{1}{2}g^{\alpha \beta}\nabla_{\alpha}\Phi^{*} \nabla_{\beta}\Phi - \frac{1}{2} m_s^2 \Phi \Phi^{*}   \right) \ ,
\end{equation}

where $R$ is the Ricci scalar associated with the metric
$g_{\alpha\beta}$, with determinant $g$,
and $m_s$ is the mass of the field. The asterisk symbol indicates the complex conjugate operation and $\nabla_{\alpha}$ denotes the covariant derivative.
Minimizing this action with respect to the metric  and with respect to the scalar field yields the Einstein-Klein-Gordon system:
\begin{equation}\label{eq:Einstein}
 R_{\alpha\beta}-\frac{1}{2}g_{\alpha\beta}R=8\pi T_{\alpha\beta} \ ,
\end{equation}
and
\begin{equation} \label{eq:KG}
g^{\alpha\beta} \nabla_{\alpha}\nabla_{\beta} \Phi = m_s^2 \Phi \ ,
\end{equation}
where $T_{\alpha\beta}$ is the stress-energy tensor 
\begin{eqnarray}
T_{\alpha\beta} &=& \frac{1}{2}\left( \nabla_{\alpha}\Phi \nabla_{\beta}\Phi^{*} + 
 \nabla_{\alpha}\Phi^{*} \nabla_{\beta}\Phi\right) \nonumber\\
 &&- \frac{1}{2}g_{\alpha\beta} 
  \left(
g^{\gamma\sigma}\nabla_{\gamma}\Phi\nabla_{\sigma}\Phi^{*} + m_{s}^2 \Phi\Phi^{*}
\right)\,.
\end{eqnarray}
In this set up, the self-gravitating scalar field interacts 
with the binary through gravity by means of the spacetime metric described by Einstein equations.

Given the total  ADM mass $M$ of a gravitational system, we define the dimensionless parameter
\begin{equation}
\tilde\mu \equiv \frac{GM m_s}{\hbar c}
\end{equation}
to characterize the scalar cloud.
This parameter is the ratio of the gravitational radius of the system $R_g = GM/c^2$, and the Compton
wavelength of the scalar field $\lambda_c = \hbar/(m_s c)$.

The linear dynamics of scalar fields propagating on a single,  non-rotating BH background has been described in  \cite{Barranco:2012qs}. It was found that regular scalar field configurations in the form of quasi-bound states around Schwarzschild BHs may survive in the vicinity of the compact objects for a certain range of values of the scalar field and BH masses. A detailed analysis of the scalar field configurations including the spectrum of quasi-bound states can be found in \cite{Barranco:2013rua}. 

The description of the scalar field assumes a harmonic time dependence,  $\Phi(t, \vec{r}) = \phi(\vec{r}) e^{-i\omega t}$,
with $\omega$ a complex number whose real part indicates the oscillating frequency and whose imaginary part determines the decay rate of the field. For small values of the dimensionless parameter $\tilde\mu$ it was found that the decay rate of the quasi-bound states decreases as a power law of $\tilde\mu$.  The mass spectrum of axion-like particles that could be probed for a given coupling $\bar{\mu}$ is continuous, and includes, among others,  
the QCD axion and a large range
of particles beyond the Standard Model predicted in the string axiverse scenario~\cite{Arvanitaki:2010}. We highlight in particular two regimes with astrophysical relevance for the combination of the scalar field and BH masses for which the
scalar field configurations may live around the central object for longer times than the age of the Universe. The first one occurs when the scalar field mass is of the order of 1 eV and the BH has a mass smaller that $10^{-17}M_{\odot}$. The second  regime corresponds to an  ultra-light scalar field with mass smaller than $10^{-22}$ eV and to a supermassive BH with mass smaller than $5 \times 10^{10}M_{\odot}$. These scenarios correspond, respectively, to axion distributions of a scalar field around primordial BHs and dark matter halos around supermassive BHs in the centers of galaxies \cite{Burt:2011pv}. For stellar size BHs as the ones employed in this work with mass $\sim 40 M_\odot$, the mass of the particle corresponds to $\sim 10^{-12}$ eV.

\section{Numerical simulations}
\label{sec:nr}

In order to write Eqs.~\eqref{eq:Einstein} and \eqref{eq:KG} as an evolution system suitable for numerical integration 
we formulate the Einstein-Klein-Gordon system using the BSSN formulation
\cite{nakamura87,Shibata95,Baumgarte98} (see also \cite{Alcubierre:2000xu, Alcubierre:2002kk}).
Our numerical simulations are performed using the open source \textsc{EinsteinToolkit} infrastructure~\cite{Loffler:2011ay,EinsteinToolkit:web,Zilhao:2013hia}. In addition, the \textsc{Carpet} package~\cite{Schnetter:2003rb,CarpetCode:web} is used for
mesh-refinement capabilities, BH apparent horizons are found using
\textsc{AHFinderDirect}~\cite{Thornburg:2003sf,Thornburg:1995cp}, and the left-hand-side of the  Einstein  equations  is  solved  using  the \textsc{MacLachlan} code~\cite{Brown:2008sb}. The scalar-field evolution code is our own modification of the the publicly available \textsc{Proca} thorn from the \textsc{Canuda} library~\cite{Canuda_2020_3565475, Zilhao:2015tya} to evolve complex scalar fields. This code has been recently employed to study the stability of spinning bosonic stars~\cite{Sanchis-Gual:2019ljs,di2020dynamical}.
The  method-of-lines with a fourth-order Runge-Kutta scheme is employed to integrate the time-dependent differential equations. 

\subsection{Initial data}
\label{subsec:sim_id}

To set initial data  suitable for numerical evolution using the “moving punctures” technique we take advantage of the Bowen-York
construction for two BHs in vacuum \cite{Bowen80}. A nontrivial analytic solution of the momentum constraint equation, thus, can be found \cite{York89}
and the Hamiltonian constraint equation can be
solved using the puncture approach \cite{Bruegmann:2006at,Hannam:2006vv, Ansorg:2004ds}. 
Once the Hamiltonian and momentum constraints are solved we introduce a nonzero scalar field distribution.
The addition of the scalar field to the binary system  introduces violations of the constraints  (see details in Appendix~\ref{app:constraint_violation}). However, our results regarding GW emission show that the initial violation produces only a weak spurious GW signal as long as the amplitude of the scalar field $\phi_0\ll1$ for $\tilde \mu \approx 1.0$. 

In our study we mainly focus on the post-merger characteristics of the gravitational waveforms. Therefore, for simplicity, we initiate our simulations when the two BHs are in their last orbit prior to merger using initial data for a quasi-circular orbit \cite{Baker_etal:2002, CookG:1994}. The BHs are positioned at $(x,y,z)=(-1.168M, 0, 0)$ and $(1.168M, 0, 0)$, having linear momentum vectors $(0, -0.333M, 0)$ and $(0, 0.333M, 0)$. 

Our simulations are performed for a non-spinning, equal mass ($m_{1} = m_{2} = 0.453 M$)  BBH system which is surrounded by a scalar field cloud initially shaped in the form of a Gaussian distribution,  $\Phi=\phi_0\,e^{-(r-r_0)^2/\lambda^2}$ centered at radius $r_0=0$, where $\phi_0$ is the initial amplitude and $\lambda$ is its half width. We simulate a series of configurations varying the amplitude $\phi_0$ between $1.0 \times 10^{-5}$ to $1.0 \times 10^{-2}$  and employing a dimensionless scalar field mass parameter ${\tilde \mu} < 1.0$. We have observed that due to the choice of constraint-violating conditions caused by the presence of the cloud numerical inaccuracies dominate the evolution of the system for values of ${\tilde \mu}$ of order $10^{-2}$. Therefore, such values are not considered. On the other hand, high-amplitude fields ($\phi_0 \approx 0.01$) trigger their collapse inside the horizon while very low-amplitude fields ($\phi_0 < 5.0 \times 10^{-4}$) lead to evolutions that are almost indistinguishable from the pure vacuum BBH case. Keeping this in mind, we simulate a fiducial number of 18  configurations, setting the scalar field mass parameter ${\tilde \mu}$ in the range from $0.3$ to $0.8$ for three scalar field amplitudes, namely $\phi_0 = 3.5\times10^{-3}, 4.5\times10^{-3}$, and  $5.5\times10^{-3}$. These simulations are compared with the vacuum BBH merger case in the absence of any scalar field. 

The mass of the cloud is kept sufficiently small, compared to the total BBH mass, to ensure that the violation of the constraints does not represent a major drawback of our initial data. For rotating BHs there is a mechanism (superradiance) that allows the cloud to grow up to about 10\% of the mass of the BH.  We take this value as an upper bound for the mass of the cloud (which scales quadratically with $\phi_0$) assuming there is no other mechanism to grow the cloud. In addition we choose $\lambda = 15$ which yields a size for the scalar cloud  comparable to the physical size of the BHs. Much larger values of $\lambda$ correspond to cloud masses that would result in significant constraint violation in our numerical evolution. Much smaller values (that are still larger than the gravitational radius of the system) would require large field amplitudes to leave any noticeable imprint in the merger waveforms. The effect of varying $\lambda$, along with the contribution of the inspiral part of the GW signal in our results, will be explored in more detail in a future work.

The numerical evolutions are performed on a Cartesian grid with a domain size of $(-320 M,\,320M)$ for all three dimensions. However, we apply reflection symmetry in the $z$ direction; thus, the computational domain in that direction is $(0, 320 M)$. The numerical grid has 9 refinement levels, starting with two centers located at each puncture, and with resolution $\lbrace$(320, 160, 80, 40, 20, 5, 2.5, 1.25, 0.625)$M$, (8, 4, 2, 1, 0.5, 0.25, 0.125, 0.0625)$M\rbrace$. The first set of numbers indicates the spatial domain of each level and the second set indicates the resolution.


\subsection{Results of the numerical evolutions}
\label{sim_bbhsc}

\begin{figure}[h!]
\includegraphics[width=0.5\textwidth]{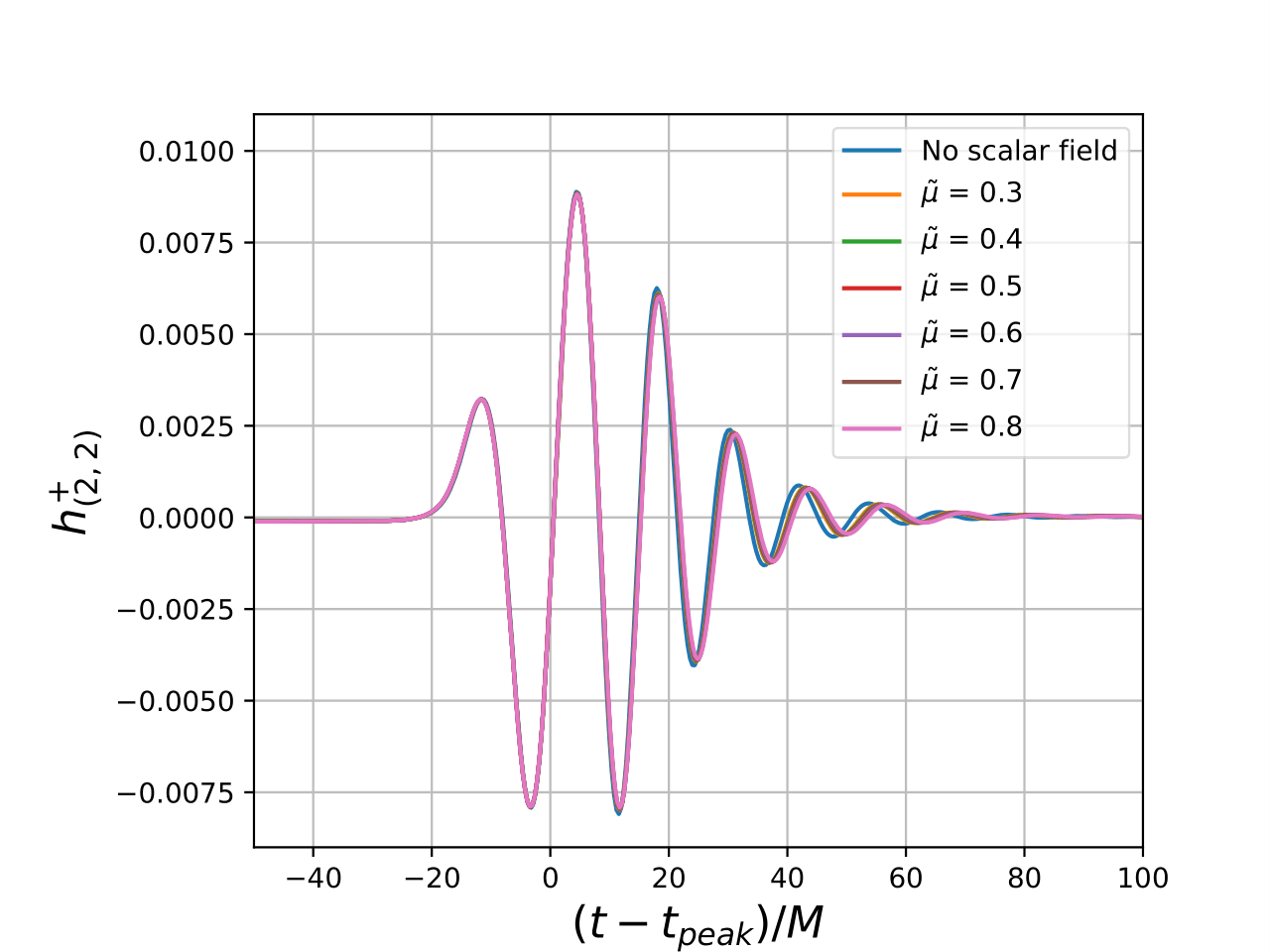}\\

\includegraphics[width=0.5\textwidth]{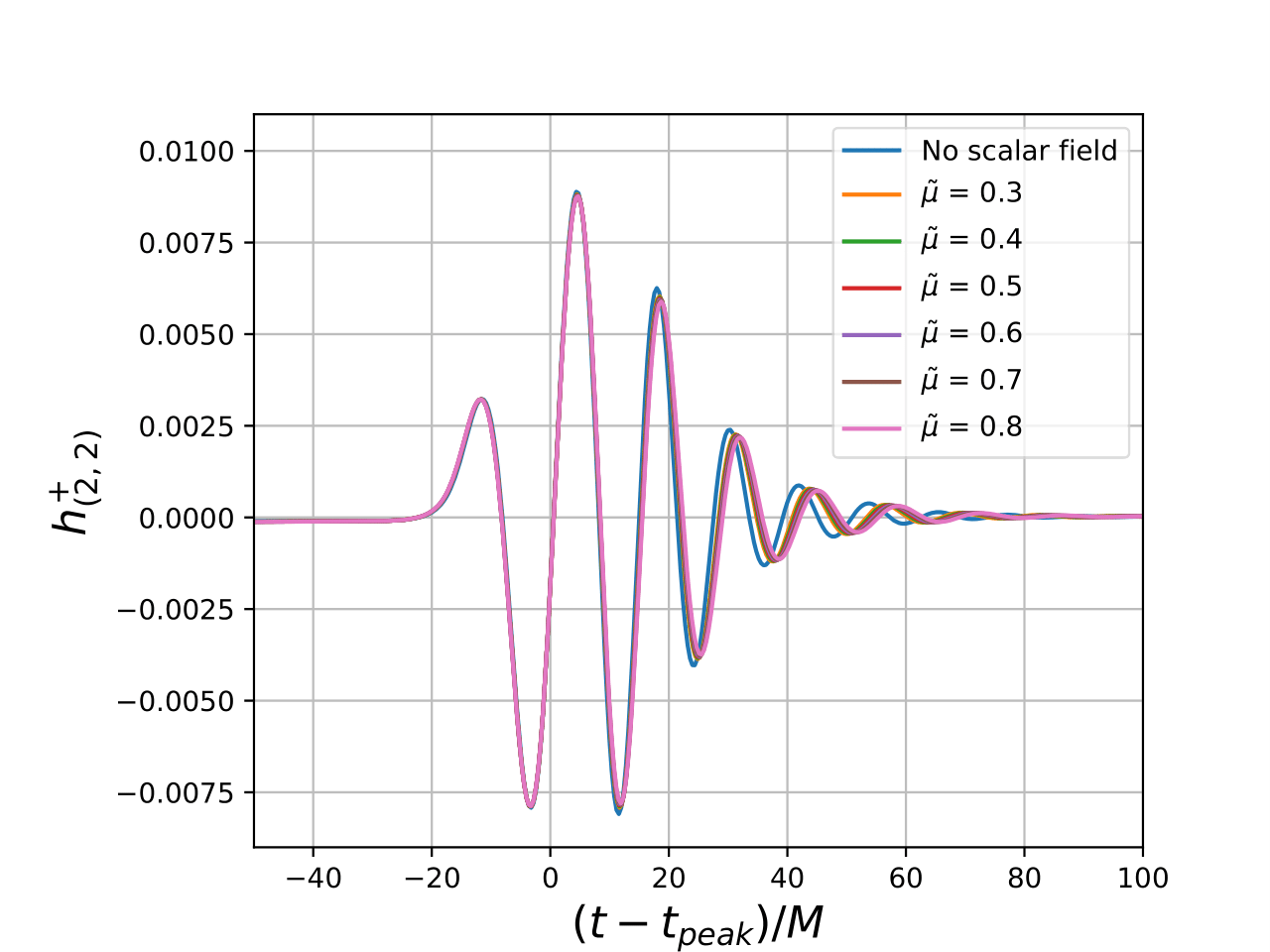}\\

\includegraphics[width=0.5\textwidth]{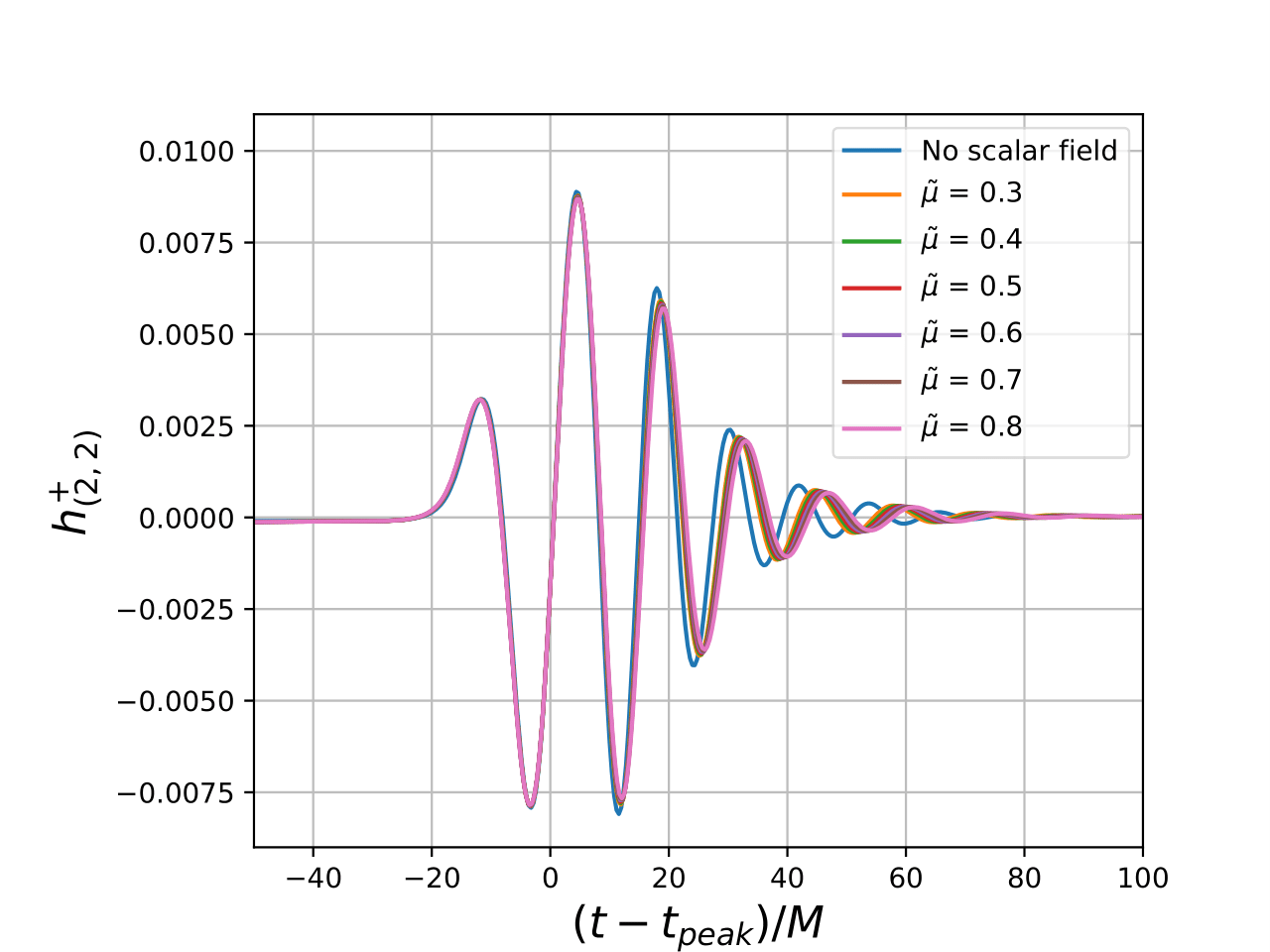}

\caption{The $l=m=2$ mode of the real part of the GW strain $h_{+}$ extracted at $r_{\rm ext}=40M$. 
Different curves correspond to varying values of the scalar field mass parameter $\tilde \mu$ for the field amplitude $\phi_0=3.5\times10^{-3}$ (top panel), $\phi_0=4.5\times 10^{-3}$ (middle panel), and $\phi_0=5.5\times 10^{-3}$ (bottom panel).
}\label{fig:h22}
\end{figure}

\begin{figure}[t]
\includegraphics[width=0.5\textwidth]{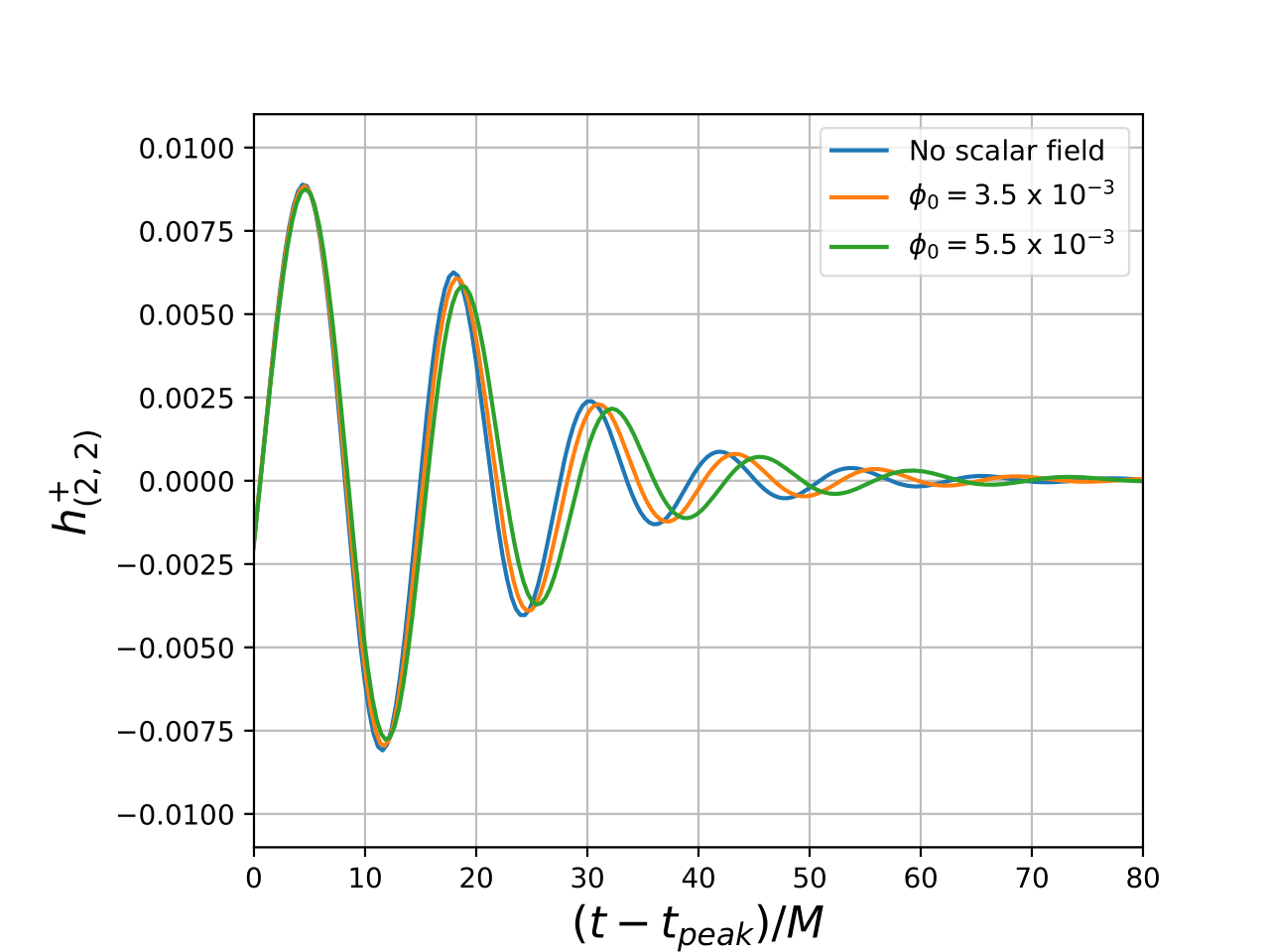}

\caption{Comparison of the ringdown quadrupolar GW strains of BBH mergers with and without scalar field content. The former corresponds to  the $\tilde \mu = 0.5$ case with $\phi_0=3.5\times10^{-3}$  and $5.5\times 10^{-3}$. The labels here are the same as in Fig.~\ref{fig:h22}.} 
\label{fig:h22_fixed_mu}
\end{figure}

Our goal is to highlight possible imprints of the presence of scalar field clouds in the gravitational waveforms produced during a BBH merger. GW signals are obtained from the simulations by computing  the Newman-Penrose scalar $\Psi_4$ defined in terms of the Weyl tensor $C_{\alpha\beta\gamma\delta}$ \cite{Newman_Penrose:1962} as
\begin{equation}
\Psi_{4} = C_{\alpha\beta\gamma\delta} k^{\alpha}\bar m^{\beta}k^{\gamma}\bar m^{\delta} \ ,
\end{equation}
where $k$ and $\bar m$ are two components of the null tetrad $l$, $k$, $m$, $\bar m$, that satisfy $\ell\cdot k=-1$ and $m\cdot \bar m = 1$, and all other inner products are zero. At a given extraction radius $r_{\rm ext}$ we perform a multipolar decomposition by projecting $\Psi_4$ onto a spherical harmonic basis of spin weight $s=-2$ as
\begin{equation}
\Psi_4(t,r,\theta,\varphi) = \sum _{\ell,m} \psi^{\ell, m}_4(t,r) {}_{-2}Y_{\ell,m}(\theta,\varphi) 
\end{equation}
whose relation with the second time derivative of the two polarizations of the GW strain is given by 
\begin{equation}
\psi^{\ell, m}_4(t,r) =  \ddot h_{\ell,m}^{+} - i \ddot h_{\ell,m}^{\times}\,.    
\end{equation}

We use the post-processing python package \textsc{pyGWAnalysis}~\cite{ETK_postprocess:web} to convert $\Psi_4$ data to GW strain.
Figure~\ref{fig:h22} displays the real part of the dominant quadrupolar ($l=m=2$) mode of the GW strain ($h^{+}$) for varying values of the mass of the scalar field and of its initial amplitude. The signal is extracted at $r_{\rm ext}=40M$ and 
$t_{\rm peak}$ refers to the instant of time when the norm of the strain waveform, $|h_{2,2}|$, reaches its maximum.
This figure shows that the presence of the scalar field produces a shift in the signal compared to the vacuum BBH case. This shift is most visible in the ringdown part of the signal and becomes larger the larger the values of $\tilde{\mu}$ and $\phi_0$.

The stronger and faster damping observed during the ringdown in the presence of high-amplitude scalar fields is highlighted in Figure \ref{fig:h22_fixed_mu}. This figure  compares the waveforms for $\phi_0 = 3.5\times 10^{-3}$ and $5.5\times10^{-3}$, both for $\tilde{\mu} = 0.5$, with a BBH merger with no scalar field content. In order to quantify this effect and to study the distinguishability of $\phi_0$ in actual GW observations, we carry out Bayesian inference with our waveform models. This is discussed in the next section.

\section{Measuring scalar field parameters in observations of BBH mergers}
\label{sec:parma_estimates}

With multiple BBH merger detections in the past and several tens to hundred expected in ground-based detectors in the coming years, it will likely become possible to distinguish BBHs with sufficiently large scalar field amplitudes from those without any such field -- or at least constrain the presence of such fields in BBH mergers. To estimate how precisely one will be able to do so, we fitted several models to the post-merger parts of our numerical-relativity waveforms, out of which the chirp sine-Gaussian waveform model came out to be the most suitable one -- partly motivated by the exponentially damped sinusoid nature of the signal in the absence of a scalar field (see, e.g.,~\cite{Aasi:2014bqj,Talukder:2014eba} and references therein). This is due to both its simplicity in structure and the small number of parameters it employs, as well as to its very high match ($\gtrsim 95\%$) with the numerical waveforms. 
The chirp sine-Gaussian form is described by the GW strain
\begin{equation}
g(t; f_0, Q, \kappa) \equiv A e^{-4\pi^2 f_0^2 t^2/ Q^2} \cos(2\pi f_0 t +\kappa t^2) \,,
\label{eqn:model}
\end{equation}
where $Q$ is the quality factor that  dictates the damping time, $f_0$ the central frequency of the sinusoid, and $\kappa$ the ``chirp parameter" quantifying the rate of change of frequency with time (see \cite{PhysRevD.94.122004} for details).

\subsection{Bayesian parameter estimation}
\label{subsec:bayesian}
\begin{figure*}
\includegraphics[width=1.0 \textwidth]{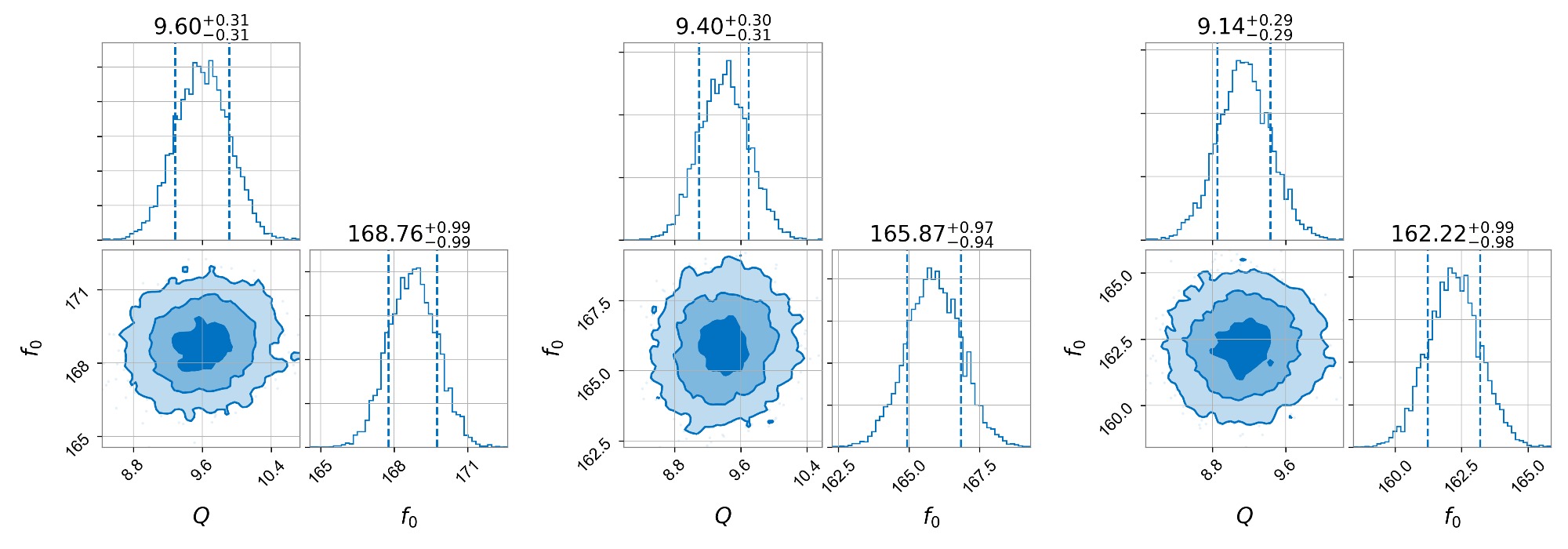}
\caption{Corner plots showing the posterior distributions of $f_0$ and $Q$ of our chirp sine-Gaussian waveform model for  $\phi_0=3.5\times10^{-3}$ (left),  $\phi_0=4.5\times10^{-3}$ (middle), and  $\phi_0=5.5\times10^{-3}$ (right). All three cases correspond to  $\tilde{\mu}= 0.50$.}
\label{fig:corner_plot}
\end{figure*}

To assess how well the  analytical model of our ringdown signals in Eq.~(\ref{eqn:model}) matches the numerical waveforms, and the precision as well as accuracy with which one can measure the BBH and scalar field parameters, we use  Bayesian inference. 
The code employed to implement it on our simulated detector data and signals is 
\textsc{Bilby}~\cite{2019ApJS..241...27A},
which is primarily designed for inferring the parameters of compact binary coalescence signals. It provides both Nested sampler and Markov Chain Monte Carlo (MCMC) sampler options for computing the parameter posteriors of modeled waveforms. In our study we used  the nested sampler ``Dynesty"~\cite{2020MNRAS.493.3132S,2004AIPC..735..395S} since it is a more appropriate choice when only a few parameters characterize a large set of waveforms. 

To perform this Bayesian analysis, the numerical waveforms are scaled such that for each set of parameter error estimates, the source is always kept at a fixed distance, namely at 450 Mpc.
Increasing the source distance leads to an increase in parameter errors that can make it difficult to distinguish the 
scalar-field values simulated here. 

We use the \textsc{IMRPhenomD} waveform templates \cite{PhysRevD.93.044006, PhysRevD.93.044007} 
with lower-frequency cut-off set high enough in our analysis so as to utilize only the post-merger parts of the signals for computing the signal-to-noise ratios (SNRs) and the parameter estimates. In particular, parameter estimation is performed using 
the quadrupolar ($l=m=2$) mode of the GW strain waveform  
as described in Section~\ref{sim_bbhsc}.  
We truncate every waveform such that only the cycles following the peak amplitude are retained.
All waveforms are injected in simulated zero-mean, colored Gaussian noise using the aLIGO's zero-detuned-high-power noise power spectral density (PSD)~\cite{alzdhp_psd}. The time-axis of the numerical waveform was scaled by setting up the component masses such that the signal frequency lies in the aLIGO sensitivity band. 
The component masses that we need for this purpose are of the order of 40$M_{\odot}$ and the mass of the scalar particle is in the range  between 
$0.5\times10^{-12}$ eV and $1.4\times10^{-12}$ eV.
Further details about the procedure are provided in  Appendix~\ref{app:match}.

\begin{table}[t]
\centering
\begin{tabular}{|p{3.25cm}|p{2cm}|p{2cm}|}
 \hline
 \multicolumn{3}{|c|}{Best Parameter fits for 100 Mpc} \\
 \hline
 $\phi_0$ and ${\tilde \mu}$ & Q & $f_0$ (Hz)\\
 \hline
 0; 0 ~\textrm{(no scalar field)} &$ 9.89_{-32}^{+34}$  & $173.54^{ +0.98}_{ -0.94}$\\
  \hline
Case $\phi_0=3.5\times10^{-3}$ \\ 
 \hline
 0.30 & $ 9.66^{+ 0.32}_{- 0.32}$  & $169.76^{+ 0.96}_{- 0.99}$ \\ 
 0.40 & $ 9.62^{+ 0.31}_{- 0.31}$  & $169.25^{+ 0.93}_{- 0.94}$ \\ 
 0.50 & $ 9.60^{+ 0.31}_{- 0.31}$  & $168.76^{+ 0.99}_{- 0.99}$ \\ 
 0.60 & $ 9.56^{+ 0.30}_{- 0.30}$  & $168.28^{+ 0.96}_{- 0.95}$ \\ 
 0.70 & $ 9.53^{+ 0.30}_{- 0.31}$  & $167.83^{+ 0.94}_{- 1.00}$ \\ 
 0.80 & $ 9.51^{+ 0.30}_{- 0.32}$  & $167.21^{+ 0.96}_{- 1.01}$ \\ 

\hline
Case $\phi_0=4.5\times10^{-3}$\\ 
\hline
 0.30 & $ 9.51^{+ 0.31}_{- 0.30}$  & $167.39^{+ 0.94}_{- 0.95}$ \\ 
 0.40 & $ 9.47^{+ 0.31}_{- 0.31}$  & $166.52^{+ 0.95}_{- 0.92}$ \\ 
 0.50 & $ 9.40^{+ 0.30}_{- 0.30}$  & $165.87^{+ 0.97}_{- 0.94}$ \\ 
 0.60 & $ 9.34^{+ 0.31}_{- 0.30}$  & $165.06^{+ 0.95}_{- 0.96}$ \\ 
 0.70 & $ 9.28^{+ 0.31}_{- 0.29}$  & $164.24^{+ 0.99}_{- 0.96}$ \\ 
 0.80 & $ 9.22^{+ 0.28}_{- 0.29}$  & $163.35^{+ 0.93}_{- 0.98}$ \\ 
\hline
Case $\phi_0=5.5\times10^{-3}$\\
\hline
 0.30 & $ 9.32^{+ 0.29}_{- 0.31}$  & $164.43^{+ 0.92}_{- 0.96}$ \\
 0.40 & $ 9.23^{+ 0.28}_{- 0.30}$  & $163.26^{+ 0.96}_{- 1.02}$ \\
 0.50 & $ 9.14^{+ 0.29}_{- 0.29}$  & $162.22^{+ 0.99}_{- 0.98}$ \\
 0.60 & $ 9.06^{+ 0.28}_{- 0.29}$  & $161.16^{+ 0.96}_{- 0.99}$ \\
 0.70 & $ 8.99^{+ 0.28}_{- 0.30}$  & $159.87^{+ 0.97}_{- 0.94}$ \\
 0.80 & $ 8.89^{+ 0.28}_{- 0.28}$  & $158.48^{+ 0.96}_{- 0.95}$ \\
 \hline

 \hline
 \end{tabular}
\caption{Variation in medians and 90\% errors in $Q$ and $f_0$ with $\phi_0$ and $\tilde{\mu}$. Here $\tilde{\mu}$ varies from 0.3 to 0.8 for three values of the scalar field: $\phi_0=3.5\times10^{-3}, 4.5\times10^{-3}$ and $5.5\times10^{-3}$. The error bars are for a source distance of 100 Mpc.  These variations are plotted in figures  \ref{fig:f0_100_450Mpc} (left panel) and \ref{fig:Q_100Mpc}.}
\label{table:final}
\end{table}

\begin{table}[t]
\centering
\begin{tabular}{|p{3.25cm}|p{4cm}|}
 \hline
 \multicolumn{2}{|c|}{Best Parameter fits for 450Mpc } \\
 \hline
 $\phi_0$ and ${\tilde \mu}$ & $f_0$ (Hz)\\
 \hline
 0; 0 ~\textrm{(no scalar field)} & $174.00^{ +4.19}_{ -4.34}$\\
  \hline

Case $\phi_0=4.5\times10^{-3}$\\ 
\hline
 0.30 & $168.10^{+ 4.44}_{- 4.22}$    \\ 
 0.40 & $167.29^{+ 4.45}_{- 4.41}$    \\ 
 0.50 & $166.67^{+ 4.60}_{- 4.45}$    \\ 
 0.60 & $165.81^{+ 4.49}_{- 4.38}$    \\ 
 0.70 & $165.00^{+ 4.52}_{- 4.39}$    \\ 
 0.80 & $164.10^{+ 4.70}_{- 4.50}$    \\ 
\hline
Case $\phi_0=5.5\times10^{-3}$\\
\hline
0.30 & $165.29^{+ 4.39}_{- 4.31}$    \\
 0.40 & $164.08^{+ 4.56}_{- 4.42}$   \\
 0.50 & $163.14^{+ 4.45}_{- 4.48}$   \\
 0.60 & $162.14^{+ 4.60}_{- 4.61}$   \\
 0.70 & $160.95^{+ 4.54}_{- 4.58}$   \\
 0.80 & $159.65^{+ 4.65}_{- 4.72}$   \\
 \hline

 \hline
 \end{tabular}
\caption{Variation in medians and 90\% errors in $f_0$ with $\phi_0$ and $\tilde{\mu}$ for a source distance of 450 Mpc. These variations are plotted in the right panel of Fig.~\ref{fig:f0_100_450Mpc}. The values for parameter $Q$ are not included in the table as their error bars overlap with the no-scalar-field case.}
\label{table:final_2}
\end{table}

The priors used for the parameters 
$Q$ and $f_0$ 
are uniform, with $Q\in [6,14]$ and $f_0 \in [140,190]$~Hz
and the likelihood used is Gaussian with sigma set equal to the standard deviation of aLIGO noise.
The posteriors thus calculated give us the estimated values and the 1$\sigma$ error bars. Some illustrative corner plots of those posteriors are shown in Fig.~\ref{fig:corner_plot}. As mentioned before, we consider three scalar-field cases  
with $\tilde \mu$ ranging from $0.3$ to $0.8$. In physical units this corresponds to $0.7\times10^{-12}$ eV - $1.8\times10^{-12}$ eV for BBH with each component of the order of 30$M_{\odot}$ and to $3.4\times10^{-13}$ eV - $9\times10^{-13}$ eV for BBH with 60$M_{\odot}$ components.
 
The $\kappa$ parameter is fixed to 13000 ${\rm Hz}^2$ in all cases as we observed it does not vary much for different values of $\phi_0$ and $\tilde \mu$. It only shows a variation of $15 \%$ when set free but does not have much effect on the match (which changes by $<1\%$).
Lists of estimated parameters are shown in Tables~\ref{table:final} and \ref{table:final_2} for source distances of 100 Mpc and 450Mpc, respectively. 

\begin{figure}[t]
\includegraphics[width=0.5 \textwidth]{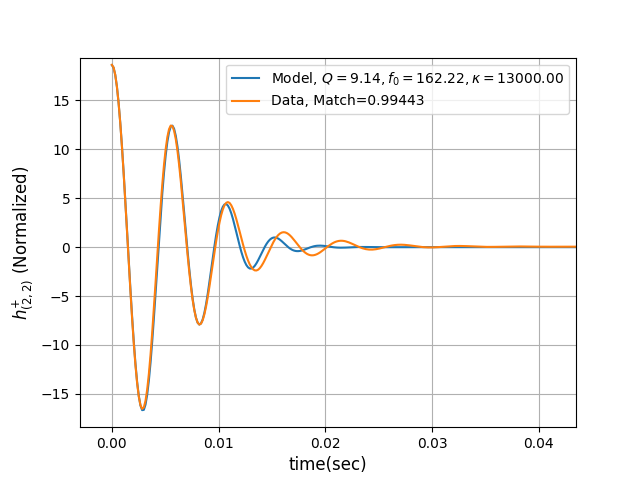}
\caption{Comparison of the post-merger waveform against the  chirp sine-Gaussian fitted model for the case $\phi_0=5.5\times10^{-3}$ and $\tilde{\mu}= 0.5$. It yields  more than 99\% match.  
}
\label{fig:Chopped waveform}
\end{figure}

 By studying injections at very high SNRs we confirmed that the
systematic error in the estimated parameters, especially, $f_0$ as listed in Table~\ref{table:final}
(where the source distance is taken at 100Mpc), is no
more than $\sim 0.1\%$.
With these estimated parameters our model gives more than 99\% match with the post merger part of the waveform, as shown in Fig.~\ref{fig:Chopped waveform} for the particular case $\phi_0=5.5\times 10^{-3}$ and $\tilde{\mu}=0.5$. 

\begin{figure}[t]
\begin{center}
\includegraphics[width=0.5 \textwidth]{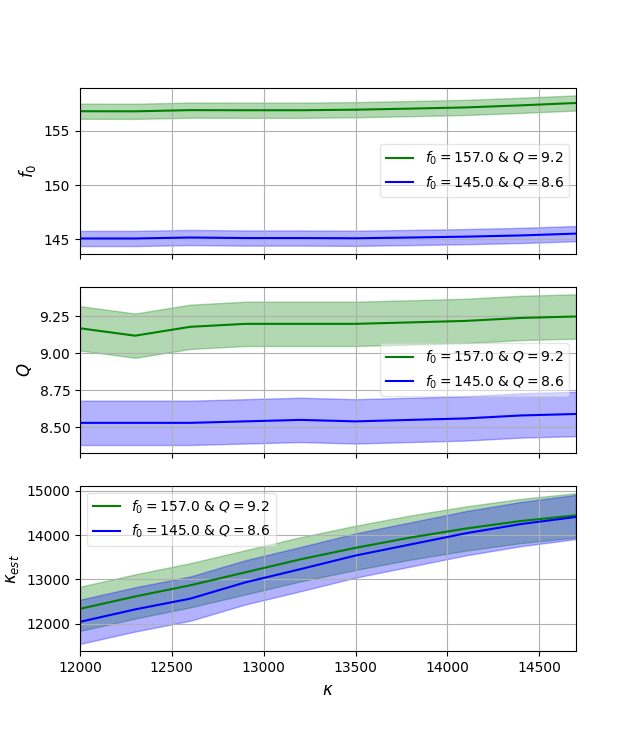}
\end{center}
\vspace{-0.5cm}
\caption{Results of our validation study for $f_0$ (top), $Q$ (middle), and  $\kappa_{\rm est}$ (bottom)
for varying values of $\kappa$.}
\label{fig:validity test}
\end{figure}

To validate our parameter estimation method we performed some tests by checking how accurate it estimated the parameters $f_0$, $Q$ and $\kappa$ of  simulated chirping sine-Gaussian signals. For this purpose, 
we varied $\kappa$ from 12000 to 14700 Hz$^{2}$ in steps of 300 Hz$^{2}$. We did it for two cases. In the first case we fixed $f_0=157.0$ Hz, $Q=9.2$ and in the second case $f_0=145.0$ Hz, $Q=8.6$. The results, shown in Fig.~\ref{fig:validity test}, demonstrate the effectiveness of our method in recovering the signal parameters.

\subsection{Revealing scalar fields properties by measuring $f_0$}
\label{subsec:phiresults}

\begin{figure*}
\includegraphics[width=1.02 \textwidth]{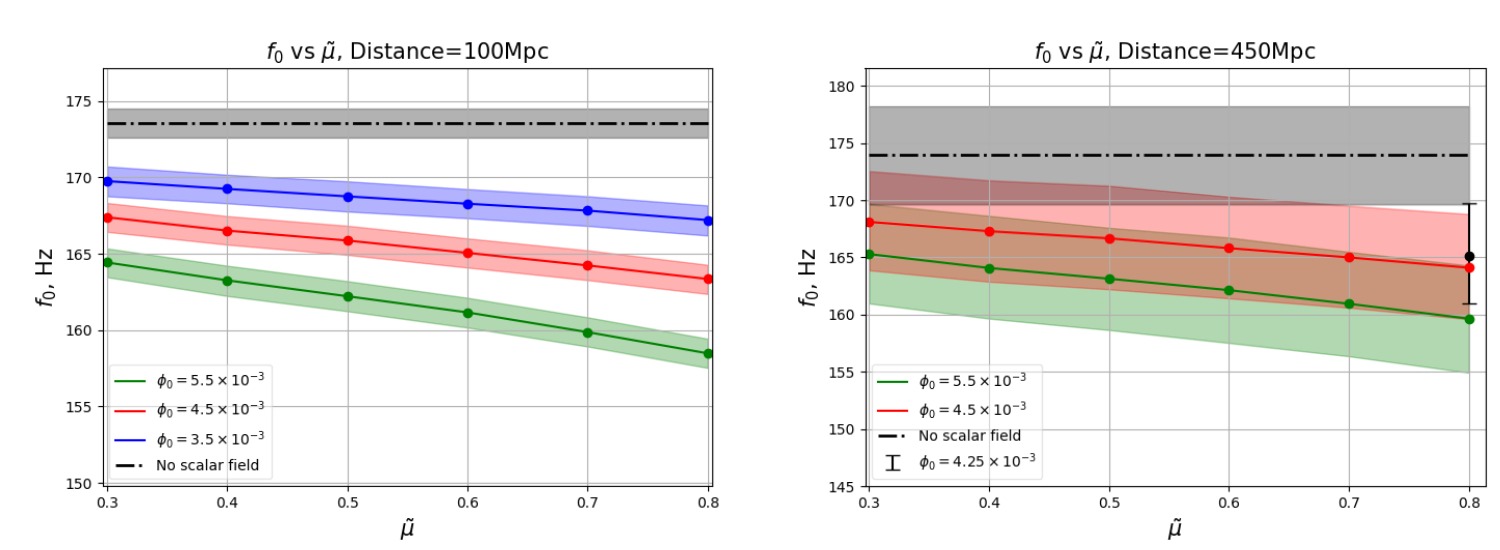}
\caption{(Left Panel) The medians and $1\sigma$ error regions (with interpolations) of the $f_0$ posteriors plotted as functions of $\tilde \mu$ for the three values of $\phi_0$ and 
for a source located at a distance of 100 Mpc. The filled circles denote the values of $\tilde \mu$ where the posteriors were individually computed. The value of $f_0$ for the no-scalar field case is $173.54_{-0.94}^{+0.98}$ Hz and it is shown as a horizontal line in the plot for reference. (Right Panel) Plot of $f_0$ vs $\tilde \mu$ for source distance of 450 Mpc. The frequency estimate for the no-scalar field case, $f_0=174.0_{4.34}^{4.19}$, cannot be distinguished from the estimate for the $\phi_0=3.5\times10^{-3}$ case for $\tilde \mu<0.7$; it can, however, be distinguished from the $\phi_0=4.5\times 10^{-3}$ case for all $\tilde \mu$ except close to $0.3$. The figure also shows that the  $\phi_0=4.25\times10^{-3}$ case is the limiting value of $\phi_0$ that can be distinguished from the no-scalar field case, close to $\tilde \mu$=0.8 
}
\label{fig:f0_100_450Mpc}
\end{figure*}

\begin{figure}[t]
\includegraphics[width=0.53 \textwidth]{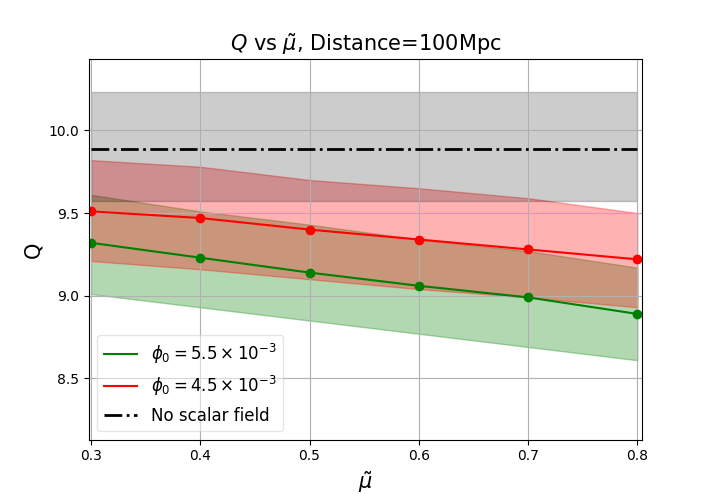}
\caption{Median and $1\sigma$ error regions of the $Q$ posteriors as a function of $\tilde \mu$ for $\phi_0=5.5\times10^{-3}$ and  $\phi_0=4.5\times10^{-3}$ and for a source at a distance of 100 Mpc. The value of $Q$ for the no scalar field case is $9.89_{-32}^{+34}$. }
\label{fig:Q_100Mpc}
\end{figure}

\begin{figure}[t]
\includegraphics[width=0.5 \textwidth]{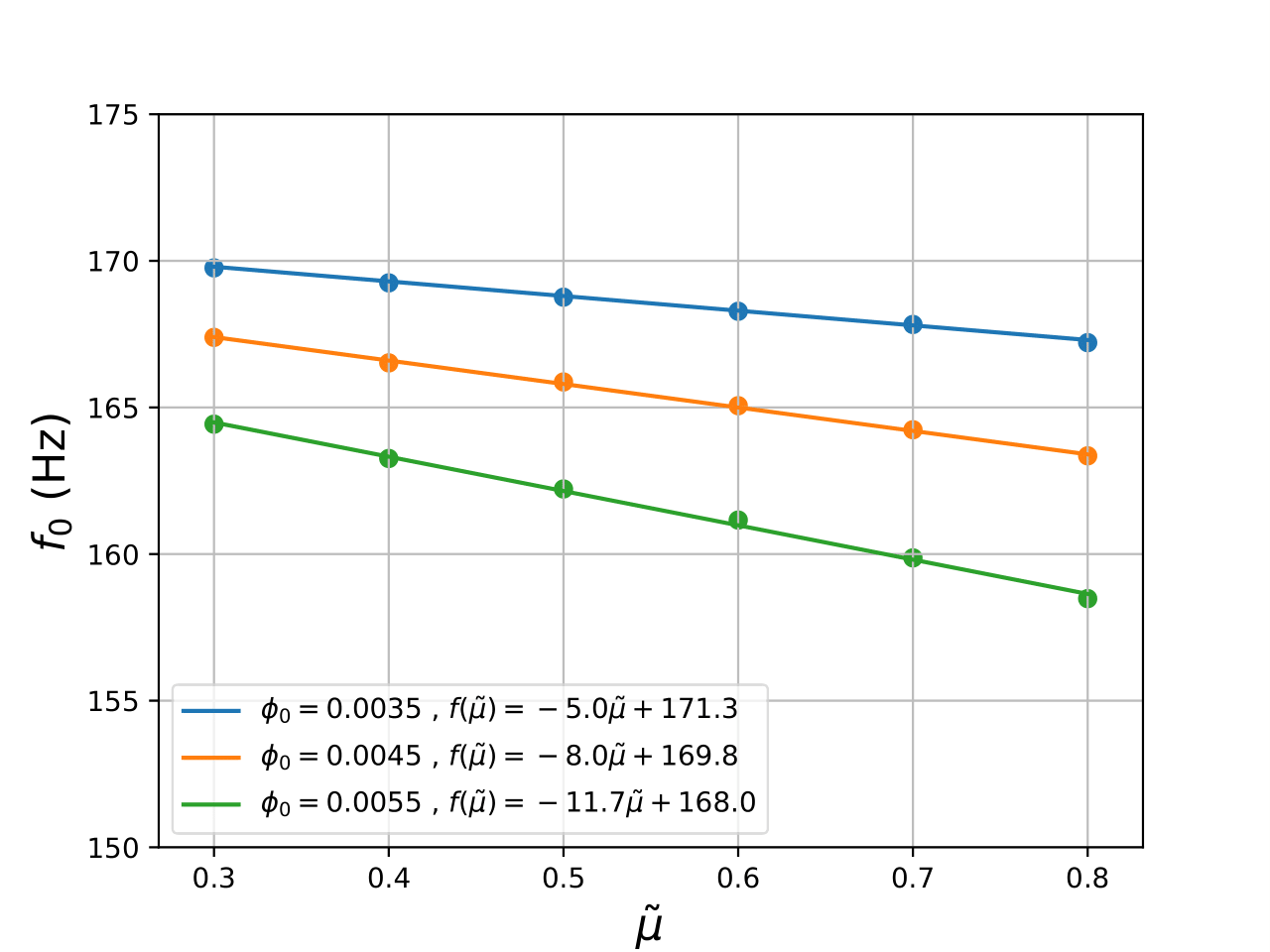}
\caption{Linear fits for $f_0$ as a function of $\tilde\mu$,  $f_0(\tilde\mu) = a \tilde\mu + b $,  for our three values of $\phi_0$.}
\label{fig:lin_fit}
\end{figure}

The signal parameter $f_0$ is an accurate tracker of the strength of the scalar field, $\phi_0$, so long as we allow for its variation with the parameter $\tilde \mu$. It is worth mentioning that for a fixed value of $\tilde \mu$, as the value of $\phi_0$ increases the mass of the scalar field cloud grows. 
We use the chirping sine-Gaussian templates in \textsc{Bilby}, as given by equation~(\ref{eqn:model}),  to measure the values of $f_0$, along with $Q$, for the multiple numerical-relativity waveforms simulated for various scalar field configurations. Specifically, we performed parameter estimation for our three values of $\phi_0$, namely, $3.5\times10^{-3}, 4.5\times10^{-3}$, and  $5.5\times10^{-3}$, as well as for our six values of $\tilde \mu$ ranging from 0.3 to 0.8. This range of values of $\phi_0$ allow us to study scalar clouds having less than 10-15$\%$ of the mass of the binary. 

Figures \ref{fig:f0_100_450Mpc} and \ref{fig:Q_100Mpc} show the variation of model parameters $f_0$ and $Q$, respectively, with the numerical waveform parameter $\tilde \mu$. The closer the  source the better the results. If the source is at 100 Mpc (left panel of Fig.~\ref{fig:f0_100_450Mpc} and Fig.~\ref{fig:Q_100Mpc}) for which the match-filtering SNR value using the post-merger part of template is $\approx$ 210, the error bars in the measurements of $Q$ are overlapping for our choices of scalar field strengths. 
However, the error bars in $f_0$ are separate and all of the 18 cases considered can be distinguished from one another, and from the no scalar-field case (with error and median values shown in Table \ref{table:final}).

\begin{figure}[t]
\includegraphics[width=0.5 \textwidth]{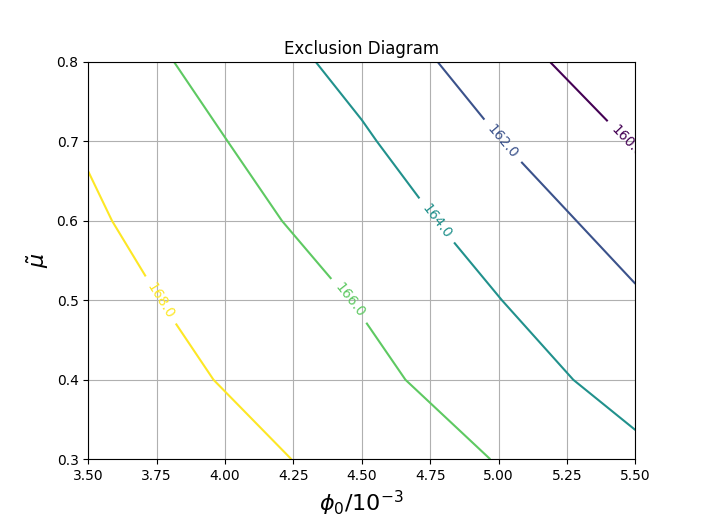}
\caption{Isocontours of $f_0$  in the $\phi_0$ - $\tilde\mu$ plane for a source at 100 Mpc.}
\label{fig:2d_contours}
\end{figure}

We note that even if the source is at a distance of 200 Mpc, one finds that the error-bars for $f_0$ remain separable for much of the $\tilde{\mu}$ range studied here. However, at larger source distances the situation worsens and it is only possible to distinguish stronger scalar fields from the no-scalar-field BBH mergers. This is shown in the right panel of  Fig.~\ref{fig:f0_100_450Mpc} for a source located at 450 Mpc for which match-filtering SNR value using post-merger part of template is $\approx$ 45, a distance around or above the values at which most GW signals from BBH coalescence events are observed by the Advanced LIGO and Advanced Virgo detector network. 

\begin{table}[t]
\centering
\begin{tabular}{|p{1.7cm}|p{1.5cm}|p{1.5cm}|}
 \hline
 \multicolumn{3}{|c|}{Model Coefficients } \\
 \hline
 $\phi_0$ & $a$ (Hz) & $b$ (Hz) \\
 \hline
0.0035 & -5.0 & 171.3  \\
0.0045 & -8.0  & 169.8  \\
0.0055 & -11.7 & 168.0 \\
 \hline
 \end{tabular}
\caption{Coefficients of fitting model for $f_0$, described by Eqs.~(\ref{eqn:fit1}) 
for the three values of the scalar field, $\phi_0= 3.5\times10^{-3}, 4.5\times10^{-3}$ and $5.5\times10^{-3}$ at 100 Mpc.} 
\label{table:fit_coeff}
\end{table}

With the previous results we can attempt to fit $f_0$ as a function of $\tilde \mu$ 
\begin{equation}
 f_0(\tilde\mu) = a \tilde\mu + b  \,.
 \label{eqn:fit1}
\end{equation}
The fits corresponding to the different values of $\phi_0$ for the source at a distance of 100 Mpc are plotted in Fig.~\ref{fig:lin_fit}. Values of the coefficients $a$ and $b$ are listed in Table~\ref{table:fit_coeff}. As can be seen from the fits (Table~\ref{table:fit_coeff}) and figures  \ref{fig:f0_100_450Mpc} and \ref{fig:lin_fit}, the frequency $f_0$ shows a clear dependence on $\phi_0$ and $\tilde\mu$ and its measurement can be used  to put bounds on the characteristics of the source.

To quantify it further we plot in Fig.~\ref{fig:2d_contours} isocontours of $f_0(\tilde\mu, \phi_0)$. An isocontour of $f_0$ specifies the region of the parameter space that is allowed by the measured value of $f_0$.  For example, if an observed GW signal has a frequency $f_0=166$ Hz, then this implies that in the range $\tilde{\mu} \in [0.3,0.8]$, $\phi_0$ must lie between $\sim 3.8\times 10^{-3}$ and
$\sim 5.0\times 10^{-3}$. Since the measured $f_0$ value will typically lie in a confidence interval, 
the range of $\phi_0$ too will have a corresponding spread. Furthermore, if the value of $Q$ can be measured as well with some precision, then along with $f_0$ it will provide a measurement of the important quantity $\tilde{\mu}$ characterizing the scalar field cloud. As suggested by Fig.~\ref{fig:Q_100Mpc} this quest might be elusive unless we detect a golden binary with a large SNR.

\section{Conclusions}
\label{sec:conclusions}

Massive scalar fields surrounding stationary and non-rotating BHs can form long-lived, quasi-bound states, or clouds, as a result of the presence of a potential well due to the mass term \citep{Barranco:2013rua}. For rotating BHs, the  superradiant instability~\citep{Brito-review} leads to the formation of hairy BHs -- Kerr BHs surrounded by bosonic (scalar or vector) hair in which the frequency of the field is synchronized with the angular velocity of the BH~\citep{herdeiro2014kerr,Herdeiro:2015gia}.
Using numerical reativity simulations we have studied mergers of BBH systems dressed in such scalar field clouds. Our aim has been to find out whether GW observations of BBH mergers could constrain the physical characteristics of a scalar field cloud surrounding those compact binaries. We have considered equal mass BBH systems endowed with Gaussian distributions of scalar field clouds parameterized by their mass $\tilde{\mu}$ and strength $\phi_0$, analyzing the imprints on the GWs generated during the mergers. We have numerically simulated the last three quarters of the final orbit prior to  merger for a large set of initial models, restricting our analysis to the post-merger phase. 

The waveforms extracted from our simulations have revealed that larger values of $\tilde{\mu}$ or $\phi_0$ cause bigger changes in the amplitude and frequency of the ringdown part of the signals.
The ringdown signals of our mergers can be simulated analytically as chirping sine-Gaussians, characterized by only three parameters, returning match values with our numerical-relativity waveforms in excess of 95\%. This is not surprising since BBH ringdown signals in General Relativity are  damped sinusoids that can be modelled with only two parameters~\cite{PhysRevD.88.122002}. 
Using our chirping sine-Gaussian signal model we have carried out computationally expensive Bayesian studies for estimating the parameters
of BBH binaries endowed with scalar field clouds. We have been able to establish that the central frequency of the model, $f_0$, has a strong dependence on the scalar-field strength $\phi_0$ and a weak dependence on $\tilde \mu$. Therefore, at a fixed value of $\tilde \mu$, a measurement of the signal parameter $f_0$ leads to a measurement of $\phi_0$. In particular, we have shown that it is possible to distinguish from observations of BBH mergers at distances of 450 Mpc, BBHs without any scalar field from those with a field strength $\phi_0 = 5.5
\times 10^{-3}$, at any fixed value of $\tilde \mu \in [0.3,0.8]$, with 90\% confidence, or better. 
We have shown that 
aLIGO may
have the potential to distinguish between the GW signal produced by a BBH
system with components of 40$M_{\odot}$ if the binary is immersed in a cloud of boson particles 
with masses between 5 $\times 10^{-13}$ eV and
1.4 $\times 10^{-12}$ eV from the BBH with the same range of masses in vacuum.

We take these results as encouraging indications for the prospect of
constraining scalar field clouds in BBH observations. However, to
assess their utility for real observations, one must study the impact of 
a wider parameter space of the binaries on how accurately and precisely one will be able to
measure the scalar field parameters from the waveforms, while also
battling possible parameter degeneracies that can arise. For instance,
the same $f_0$ and $Q$ values can correspond to a variety of BBH
remnant masses and spins as well as scalar-field parameters.
(Arguably, some of these degeneracies may break or get mitigated by measurements of source parameters in the inspiral part of the signals.)
 To address this issue, one
will need to simulate waveforms for a broader range of astrophysically relevant BBH
component masses and spins, and scalar-field parameters, and analyze the parameter degeneracies that might
arise as well as their possible resolution. We leave this
computationally expensive study for the future.

While in this investigation we have limited ourselves to single BBH observations, we note that by combining multiple BBH detections, one may be able to constrain the scalar field configurations in these mergers collectively for populations. Such an exercise will be similar to stacking ringdown signals from multiple BBH signals to, e.g., test the no-hair theorem. In our case, however, a straightforward extension to populations is complicated by the possibility that the scalar field parameters may vary from one BBH source to another. Similarly, it would be interesting to explore if BBH observations can be used to determine or constrain the mass of ultra-light bosons. We plan to pursue these prospects in a future work.

\acknowledgments

We are thankful to Cristiano Palomba  and the referee for their useful comments and suggestions. This research is supported by the Spanish Agencia Estatal de Investigaci\'on (grants PGC2018-095984-B-I00), by the Generalitat Valenciana (PROMETEO/2019/071), by the European Union’s Horizon 2020 RISE programme H2020-MSCA-RISE-2017 (Grant No.~FunFiCO-777740) and by
DGAPA-UNAM through grant IN105920.
 A. G. is
supported, in part, by the Navajbai Ratan Tata Trust research grant. NSG is supported by the Funda\c c\~ao para a Ci\^encia e a Tecnologia (FCT) projects PTDC/FIS-OUT/28407/2017, UID/FIS/00099/2020 (CENTRA), and CERN/FIS-PAR/0027/2019.
The numerical simulations for this work were performed on the Perseus and Pegasus clusters, which are part of the high performance computing (HPC) facility at The Inter-University Centre for Astronomy and Astrophysics, Pune, India (IUCAA). We would also like to acknowledge help and support of their staff.   


\appendix

\section{Constraint violation in the presence of a scalar field}
\label{app:constraint_violation}

\begin{figure} 
\includegraphics[width=0.5\textwidth]{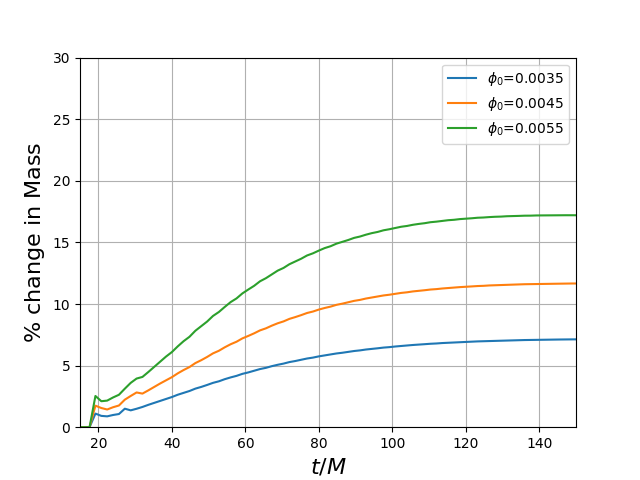}
\includegraphics[width=0.5\textwidth]{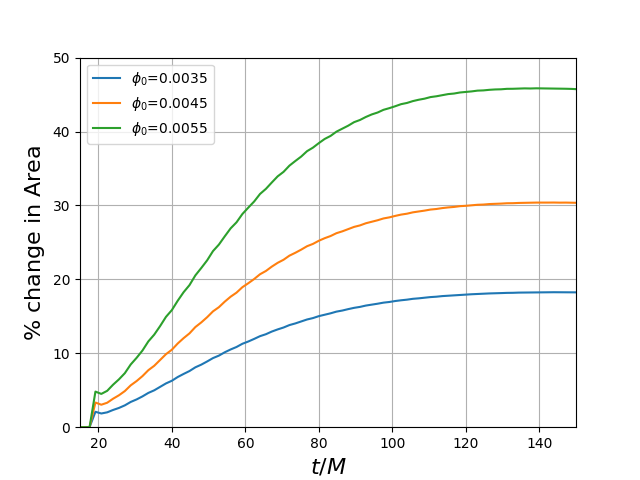}
\caption{Percentage change in mass (top panel) and area (bottom panel) with respect to the no scalar field case for the largest mass parameter considered $\tilde\mu = 0.8$ and for field strength $\phi_0=3.5\times10^{-3}$, $4.5\times 10^{-3}$ and $5.5\times 10^{-3}$.  
}\label{fig:mass_change}
\end{figure}

\begin{figure}[t]
\begin{center}
\includegraphics[width=0.5\textwidth]{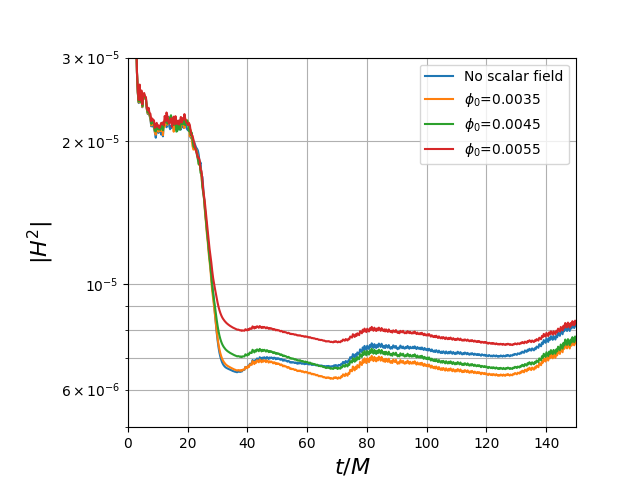}\\
\end{center}
\caption{Time evolution of the $L_2$-norm of the Hamiltonian constraint for BBH mergers with and without a scalar-field cloud. The data shown correspond to the largest mass parameter  considered in this work, $\tilde\mu = 0.8$,   for the field strengths $\phi_0=3.5\times10^{-3}$, $4.5\times 10^{-3}$,  and $5.5\times 10^{-3}$  } 
\label{fig:ham_norm2}

\end{figure}

As described in section~\ref{subsec:sim_id}, the introduction of a scalar-field distribution in an otherwise constraint-preserving BBH initial setup leads to constraint violations. To make sure that our numerical simulations are still valid for the time evolution under consideration, we compute the changes in the mass and the area of the black hole and compare those values with the no scalar field scenario. 

The top panel of Figure~\ref{fig:mass_change} shows the percentage change in mass whereas the bottom panel indicates the corresponding change in area, for the largest mass parameter considered, $\tilde\mu = 0.8$, and for three values of the field strength.

 To compute the area, mass and other physical quantities on the horizon, 
first we find the horizon using the \textsc{AHFinderDirect} thorn~\cite{Thornburg:2003sf,Thornburg:1995cp} followed by the application of the  \textsc{QuasiLocalMeasures} thorn, which implements the isolated and dynamical horizon  framework~\cite{Dreyer:2003ihnr,Ashtekar:2004lrr, Schnetter:2006dhnr}. This framework enables the computation of quasi-local quantities on marginally trapped surfaces such as the apparent horizon. 
Recently, the correlation between the shear of the horizon and the News function in the wave-zone was demonstrated  for  
quasi-circular BBH
mergers \cite{gupta_etal_prd_2018, prasad_etal_prl_2020} and horizon dynamics 
was studied 
in terms of the shear and the multipole moments for the head-on collision of two black holes \cite{Pierre-etal:2020c}.  These works could be important steps towards inferring finer details of black hole horizon properties through gravitational-wave observations. It might be interesting to pursue similar studies in the presence of scalar field clouds, 
which we aim to do in the future.

In Fig.~\ref{fig:ham_norm2} we show the evolution of the $L_2$-norm of the Hamiltonian constraint for the simulations of Fig.~\ref{fig:mass_change}. 
We find that the magnitude of the violation of the $L_2$-norm of the Hamiltonian constraint is comparable to the no scalar field case for the range of field strengths $\phi_0$ and mass parameter $\tilde\mu$ considered. We note that since these results correspond to the largest mass parameter of our study, the impact of the constraint violation is in general much lower for the $\phi_0$-$\tilde\mu$ parameter space investigated.

\section{Pre-parameter estimation treatment of numerical data} 
\label{app:match}

Since the numerical waveforms are computed here in terms of $\psi_4$,  as functions of time in code units of the Einstein toolkit~\cite{EinsteinToolkit:web}, it is necessary to convert the $\psi_4$ data into GW strain data and convert the code time units into seconds (physical units) so that they are usable for GW observations and parameter measurement projections.
Here we show how such a conversion is done.
The conversion from code units to seconds depends on the total mass of the system $M_{\rm total}$. This fact can be used to adjust the BBH mass so as to bring the frequency parameter of the numerical waveforms into aLIGO's most sensitive region band, namely, 100-200~Hz.

Moreover, Some care must be borne when computing the match with numerical relativity (NR) waveforms, as we explain below:

\begin{enumerate}
\item  The NR simulations produce $\psi_4$ data for our various scalar field configurations.
Therefore, for GW data analysis, we first construct 
GW strain waveforms from these data.

\item  For labeling the time-points of the strain data in physical units, we used the following conversion 
\begin{equation}
    t({\rm in~sec})=M_{\rm total}\times \frac{M_\odot G}{c^3}*({\rm cctk\_time})\,
    \label{eq:time_conversion}
\end{equation}
 where $\rm cctk\_time$ refers to the code time units in the Einstein Toolkit~\cite{EinsteinToolkit:web}.

\item Each  waveform strain time-series was chopped, resampled and zero padded to prepare it for our analysis with only the post-merger piece (i.e.~the part of the NR waveform that starts at the time-point where the peak amplitude is attained in the time-domain); see 
Fig.~\ref{fig:Chopped waveform}.

\item As we wanted our waveforms in the aLIGO sensitivity band, we chose the masses such that the frequency parameter falls in that region. To get the approximated value of those masses we calculated the match of numerical waveforms with the IMRPhenomD~\cite{PhysRevD.93.044006,PhysRevD.93.044007} template in the region (100,200)Hz [the approx aLIGO sensitivity band]. The match comes out to be very high in the mass region of $30-60M_{\odot}$. On this bases we chose component masses to be $40M_{\odot}$.
\item We also calculated frequency of the numerical waveforms by measuring the half cycle and concluded that we need around $40-40M_{\odot}$ BBH to get our frequency in aLIGO sensitivity band. 

\item In our study we analysed  numerical waveforms at different source distances. This we achieve by making the peak amplitude of numerical waveforms equal to the PyCBC\cite{alex_nitz_2020_3969565} generated IMRPhenomD template for the distance we want to study.

\end{enumerate}

\bibliography{bbh_scloud.bib}
\end{document}